\documentclass[reprint, pra, aps, superscriptaddress]{revtex4-2}

\usepackage[utf8]{inputenc}
\usepackage{graphicx}
\usepackage{dcolumn}
\usepackage{bm}
\usepackage{amsmath, amssymb}
\usepackage{xcolor}

\begin{document}

\title{Electron-positron pair generation using a single kJ-class laser pulse in a foam-reflector setup}

\author{Oliver Mathiak}
\email{oliver.mathiak@hhu.de}
\affiliation{Institut f\"{u}r Theoretische Physik I, Heinrich-Heine-Universit\"{a}t D\"{u}sseldorf, 40225 D\"{u}sseldorf, Germany}

\author{Lars Reichwein}
\affiliation{Peter Grünberg Institut (PGI-6), Forschungszentrum Jülich, 52425 Jülich, Germany}
\affiliation{Institut f\"{u}r Theoretische Physik I, Heinrich-Heine-Universit\"{a}t D\"{u}sseldorf, 40225 D\"{u}sseldorf, Germany}

\author{Alexander Pukhov}%
\affiliation{Institut f\"{u}r Theoretische Physik I, Heinrich-Heine-Universit\"{a}t D\"{u}sseldorf, 40225 D\"{u}sseldorf, Germany}

\date{\today}
\begin{abstract}
    We investigate the process of creating electron-positron pairs from laser-matter interaction in pre-ionised foam targets using particle-in-cell simulations. A high-intensity laser pulse drives electrons via direct laser acceleration up to a cone-shaped reflector. The high-energy electrons interact with the reflected laser pulse, generating abundant pairs. The effects of the plasma-channel shape on the propagation of the laser pulse and subsequent pair production is studied. The results show that the number of Compton emission and Breit-Wheeler pair creation events is highly sensitive to the diffraction of the laser due to its interaction with the foam.
\end{abstract}
\maketitle

\section{Introduction}
The creation of matter and antimatter from light is one of the most prominent predictions of quantum electrodynamics (QED). Although this effect has been experimentally shown in the presence of nuclei \cite{Blackett1933, PhysRev.50.263,PhysRev.43.491,Chen2009-lr}, the creation of pairs from the collision of two or more light quanta remains elusive. 
Numerous theoretical and experimental setups have been proposed to generate a large number of pairs with contemporary or near-future laser facilities \cite{Blackburn_2018,LUXE_Collaboration2024-fm,PhysRevA.107.012215, Filipovic2022-yl, Samsonov2025-kf}, but due to the high energy required, generating a large number of pairs remains challenging. This makes studying the collective behaviour of electron-positron plasmas challenging, as they require an especially large number of pairs \cite{Mathiak2025-rr, 6341853,osti_4723126}.
Such plasmas feature prominently in some of the most energetic astrophysical phenomena, including gamma ray bursts and pulsar winds, and represent a unique state of matter \cite{Rylov1981-ew, Iwamoto1989-ne,Iwamoto2002-wk,Takahara1986-te,Chen2023-yb}.

With the advent of ultra-intense lasers, such as the 10 PW L4 laser at the ELI facility \cite{Condamine2023-ib}, the experimental realisation of $e^-e^+$ plasmas is finally coming within reach. With these systems, it has become possible to probe new frontiers of high-energy density laser-plasma interactions, where strong-field quantum electrodynamic (SF-QED) effects not only become relevant, but dominant. As laser intensities approach the critical Sauter-Schwinger limit $E_\mathrm{cr} = m^2c^3/e\hbar$ corresponding to an intensity $\sim 10^{23} \, \mathrm{W/cm}^2$, the classical description of the interaction of particles with the field is no longer sufficient. Instead, a full quantum-mechanical description is needed. Such descriptions have been implemented in state-of-the-art particle-in-cell (PIC) codes as probabilistic Monte-Carlo methods \cite{Fedeli2022-cx,Montefiori2023-vi}, enabling the simulation of QED effect in laser-plasma interactions.

The generation of electron-positron pairs is facilitated by the non-linear Breit-Wheeler process, where a single high-energy photon collides with multiple field-photons to generate a pair. Generating the required high-energy photons in a sufficient number can be achieved by bremsstrahlung in high-Z targets \cite{Sarri2015-pf, Chen2009-lr} or by collision of a high-energy electron with multiple background photons, namely, nonlinear inverse Compton scattering. This occurs commonly in high-intensity laser-plasma interactions, making it a natural source for high-energy photons in this context.

For efficient pair creation,  a large number of high-energy electrons is required. Among the established acceleration mechanisms in laser–plasma interactions, laser-driven wakefield acceleration (LWFA) can produce GeV-scale electrons over an acceleration distance of a few centimetres \cite{Picksley2024_10GeV}. However, the maximum attainable beam charge is limited (typically to a few pC) depending on the specific regime \cite{Jiang2022-ns}. In contrast, direct laser acceleration (DLA) can provide energetic electrons in much larger numbers (multiple nC) \cite{Babjak2024-zp}. While LWFA is often preferred as it is able to produce quasi-monoenergetic particle beams, this is not a requirement for the process of pair creation, making DLA a more suitable acceleration method.

DLA is dominant for targets of near-critical density $n \sim 0.1-1 n_c$ ($n_c$ being the critical density), where the laser pulse can efficiently accelerate the electrons without dephasing \cite{Pukhov1999-gd,Gahn1999-qq,Babjak2024-zp}. Foam targets are well-suited to produce such plasmas and have therefore garnered increasing interest for the study of high-energy laser plasma interactions \cite{PUGACHEV201688,Rosmej,Gyrdymov2024-nd}. Their defining porous structure renders them largely optically transparent to incident laser pulses, allowing an ionising laser pulse to travel through the target undisturbed and rapidly convert it from a structured solid to a nearly homogeneous plasma \cite{Tikhonchuk, Nagai2018-ew}. While gas jet targets are only capable of producing low-density plasmas (typically $\lesssim 0.1 n_c$ \cite{Gilljohann2019-od, Sarri2015-so}), foam targets can generate plasmas with near-critical or even overcritical densities ($\gtrsim   0.6 n_c$) \cite{Tikhonchuk, Rosmej2025-rc, Mariscal2021-sp}. As such, they can bridge the gap between gas jet targets and solid-state targets. Such properties are advantageous for many applications such as inertial confinement fusion, bright x-ray sources, laser shock measurements, and laboratory astrophysics, among others \cite{Tikhonchuk}.

In this paper, we propose a single-laser setup, where a high-intensity laser pulse acts as both the accelerator of electrons via direct laser acceleration and as a colliding laser pulse to generate a large number of electron-positron pairs. We consider a pre-ionised, homogenised hydrogen foam (in which the electrons get accelerated) with a cone-shaped gold reflector target placed behind it. Using 3D-PIC simulations, we investigate the acceleration of electrons inside the foam plasma, their DLA-like behaviour and resulting energy spectra. Using higher-resolution 2D simulations, we further study the dependence of pair yield on different plasma parameters. We conduct a qualitative comparison of our results with known theory.

In section \ref{sec:theory}, we describe the theoretical basis of the SF-QED effects that we consider in our simulations. The simulation setup is detailed in \ref{sec:setup}, including a discussion of our sub-sampling routine for QED effects. Finally, the results and discussion are presented in \ref{sec:results}.

\begin{figure}
    \centering
    \includegraphics[width=1\linewidth]{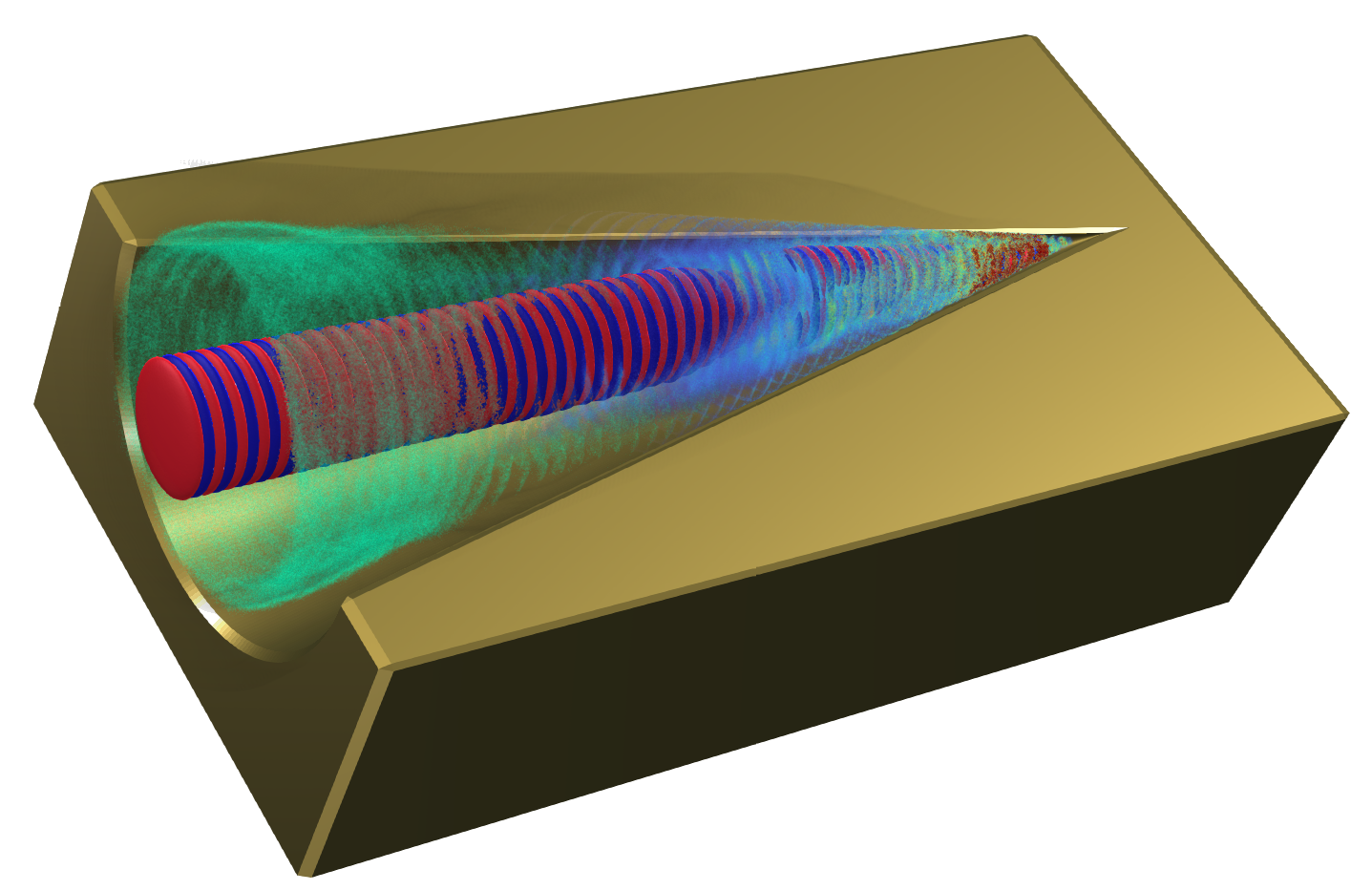}
    \caption{Visualisation of the laser pulse propagating through the pre-ionised foam inside a cone-shaped cavity within a solid reflector target. The laser expels electrons transversely, forming a channel structure (green). Once the laser pulse is reflected at the cone target, a region with high $\chi$-parameter is generated (blue).}
    \label{fig:3d}
\end{figure}

\section{SF-QED Effects} \label{sec:theory}

The creation of electron-positron pairs by the interaction of high-energy electrons with strong laser fields is facilitated by a two-step process. Electrons can interact with background fields, emitting hard photons. Those in turn can interact with the background fields to decay into an electron-positron pair. The dimensionless quantum nonlinearity parameter
\begin{align}
    \chi &= \frac{e \hbar}{m_e^3 c^4} \sqrt{(F_{\mu\nu} p^\nu)^2} \notag \\
    &= \frac{\gamma}{E_\mathrm{cr}} \sqrt{\left(
        \mathbf{E + \mathbf{v} \times \mathbf{B}}\right)^2 -\left( \mathbf{E} \cdot \mathbf{v}/c \right)^2}  \label{eq:chi}
\end{align}
determines the strength of the processes. Here, $e$ is the electron charge, $\hbar$ the reduced Planck constant, $m_e$ the rest mass of the electron, $c$ the speed of light, $F_{\mu\nu}$ the electromagnetic field tensor, and $p^\nu$ the particle's four-momentum. Furthermore, $E_\mathrm{cr} = m^2c^3/e\hbar$ is the critical Sauter-Schwinger field and $\gamma$ the Lorentz factor of the particle.

In a strong background field, a high-energy lepton can interact with multiple field (soft) photons to create a high-energy (hard) photon. The energy loss of the lepton due to the photon emission is the quantum-mechanical continuation of the classical radiation reaction \cite{Ritus1985-sq}. The total rate of photon emission can be approximated as \cite{10.1007/3-540-55250-2_37}:
\begin{align}\label{eq:rate_photon}
    W_\gamma = \frac{5}{2\sqrt{3}} \frac{\alpha \chi_\gamma}{\tau_e \gamma} \frac{1}{\sqrt{1+\chi_\gamma^{2/3}}} \, .
\end{align}

A high-energy photon in turn, can interact with multiple background photons to create an electron-positron pair. This approximation can be written in closed form as \cite{10.1007/3-540-55250-2_37}:
\begin{align}
    W_\mathrm{PP} \approx \frac{\alpha m_e^2}{\hbar\varepsilon_\gamma} \frac{0.23 \chi_\gamma}{(1+ 0.22\chi_\gamma)^{1/3}} \exp(-8/3\chi_\gamma)\,. \label{eq:rate_positron}
\end{align}
Evidently, the pair production rate is exponentially suppressed for $\chi_\gamma \ll 1$.
In the locally-constant crossed field approximation (LCFA), where we assume the fields to be orthogonal, $\mathbf{E} \perp \mathbf{B}$, and locally constant in time, $\mathbf{B}(t) = \mathbf{B}_t$, for some time step $\Delta t$, the non-linear Compton scattering and Breit-Wheeler process can be modelled as a probabilistic Monte-Carlo algorithm \cite{PhysRevE.92.023305, RIDGERS2014273}.

\section{Setup} \label{sec:setup}

An electron co-propagating with a laser pulse does not experience any quantum electrodynamical effects, because the contributions of the fields to the quantum nonlinearity parameter $\chi$ cancel each other exactly. In contrast, $\chi$ is maximised if the electron is counter-propagating with the laser pulse. Thus, common setups for pair production utilize a colliding electron or photon beam with a laser pulse, requiring a two-stage setup; a first stage where high-energy electrons or photons are generated, and a second stage where the electron-positron pairs are generated. Here, we instead propose a single-stage setup, where a single laser pulse acts both as an accelerator for the particles via DLA and as the colliding laser beam by reflecting it off a gold reflector target (cf. Fig. \ref{fig:3d}).  

Since $\chi$ depends on both the particle momentum and field amplitude, achieving simultaneously high electron energies and strong fields is essential for maximising the pair yield.
To accelerate electrons, we consider a pre-ionised near-critical hydrogen plasma. Such a plasma can be readily produced by irradiating a foam target with a moderate laser pulse, which vaporises and pre-ionises the solid structure.
We consider a plasma density profile of a Gaussian channel at the optical axis, described by
\begin{align}
        n(r) = n_0 + (n_1 - n_0) \exp(-r^2 / 2\sigma_C^2)\,,
\end{align}
with inner density $n_0$, outer density $n_1$ and channel width $\sigma_C$. Behind the foam, a solid, gold reflector target with a cone-shaped cavity and opening angle $\alpha$ reflects the laser pulse and amplifies the fields to multiples of the laser amplitude.

For the laser pulse, we consider a linearly polarised laser pulse propagating along the optical axis $\mathbf{e}_x$. It has a Gaussian longitudinal and transverse profile of length $T = 150$ fs and a focal spot size $w = 5$ µm. With a wavelength $\lambda = 800$ nm and a total pulse energy $1.5$ kJ this results in a laser parameter $a_0 = E/E_0 = eE / m_e c \omega_L = 194 $, where $\omega_L = 2\pi c/\lambda$ is the laser frequency.

We study the proposed setup utilising the fully electromagnetic 3D particle-in-cell code \textsc{vlpl} \cite{VLPL, Pukhov2016}. The QED effects (cf. section \ref{sec:theory}) are simulated using stochastic Monte-Carlo methods \cite{PhysRevE.92.023305, RIDGERS2014273}.

We use a simulation box of size $240\lambda \times 90\lambda \times 90\lambda $ with a grid resolution of $h_x = 0.1 \lambda$, $h_y = h_z = 0.1 \lambda$ and a time step $\Delta t = 0.07 \lambda / c$ with 4 particles per cell for our 3D simulations.

Realistic estimates of pair yield require fully three-dimensional simulations, because there is a significant difference in energy gain via laser acceleration and photon emission between 2D and 3D simulations \cite{PhysRevE.104.045206}. However, such simulations are enormously computationally demanding and therefore impractical for a systematic investigation of the dependence on different plasma and reflector parameters. Fortunately, while the overall energy balance between 2D and 3D simulations differs significantly, 2D simulations still allow for a qualitative investigation of the underlying behaviour. Accordingly, we conduct 2D simulations to investigate the behaviour of the pulse and the accelerated electrons in dependence on the plasma and reflector parameters.

\subsection*{Sub-sampling in PIC simulations}

The aforementioned laser pulse can accelerate resonant electrons in the channel via direct laser acceleration (DLA) to energies of up to $5$ GeV ($\gamma \sim 10^4$) \cite{Babjak2024-zp}. For our choice of field amplitude, we can approximate $\chi \gg 1$ under the assumption that the electric and magnetic fields are orthogonal to the particle momentum and each other and $v \approx c$, where Eq. \eqref{eq:chi} reduces to $\chi \approx \gamma E / E_\mathrm{cr}$.

With this, we are well into the regime where quantum effects are relevant and electron-positron pair production becomes possible. However, the bulk of the electrons are still below the threshold $\chi \lesssim 1$, where pair-production is exponentially suppressed. Therefore, the total rate of pairs created depends heavily on the field amplitude and the electron energy.

Since the rates for photon emission and pair creation differ by several orders of magnitude in the barely-quantum regime $\chi \lesssim 1$ (compare Eqs. \eqref{eq:rate_photon} and \eqref{eq:rate_positron}), generating a sufficient number of numerical pairs for decent statistics is computationally challenging. 

This is because usually, the generated pairs have the same granularity as the photons and initial electrons. Consequently, a large number of numerical photons have to be considered to get the number of numerical positrons required for good statistics. This fact can be remedied by restricting the phase space of the considered photons and by using a simple sub-sampling scheme.

The bulk of the generated photons are low-energy, meaning their probability to decay into an electron-positron pair is negligible, as the rate is exponentially suppressed for low $\chi$. At the same time, they are energetic enough not to interact with the plasma and can therefore be safely discarded. As such, we have a lower bound of energy of photons we have to consider.

Although this can alleviate much of the computational load, the granularity of the positrons is still limited by that of the initial electrons.

To get a better pair-yield resolution, we employ a simple sub-sampling scheme for our pair-production method. Numerical photons are effectively split into an arbitrary number of virtual sub-particles, perform the Monte-Carlo event for each sub-particle individually, and sum up the resulting pairs. The result follows a binomial distribution and can be simulated using a simple random sample from such a distribution. This greatly improves the resolution of the pair production process with a moderate increase in computational load.
More details on the sub-sampling routine can be found in the proceeding \cite{Mathiak2025_proc}.

\begin{figure}[h]
    \centering
    \includegraphics[width=1\linewidth]{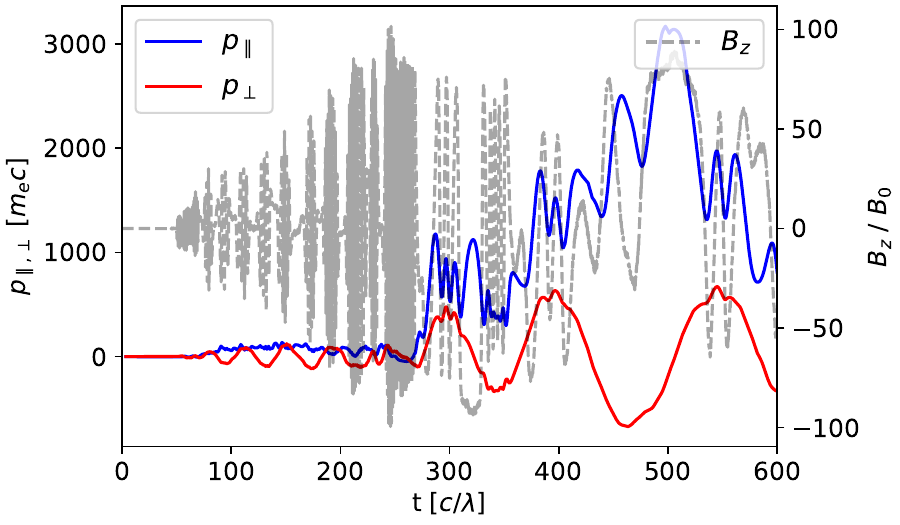}
    \caption{Longitudinal (blue) and perpendicular(red) momentum of a particle with the $B_z$ (grey) field it experiences.}
    \label{fig:traj}
\end{figure}

\section{Results and Discussion} \label{sec:results}

In low-density $n_0 \lesssim 0.1 n_c$ homogeneous plasmas, such as those typically created by gas jets, a strong laser pulse diverges quickly. Due to the high intensity of the laser pulse, electrons and ions are rapidly expelled, leaving an effectively matter-free space in which the laser pulse diverges as it would in vacuum. Due to the high intensity of the laser pulse, the plasma response is no longer sufficient to focus the laser. 
Consequently, to maintain high field strengths at the reflector target, either a very short plasma $(L \ll L_\mathrm{Rayleigh})$ or an additional focusing mechanism is required. 
Short plasmas are unsuitable at low densities because the laser pulse needs sufficient time to accelerate the electrons to high energies. This problem can be alleviated by increasing the plasma density, thereby reducing the length required to accelerate the electrons to high energies. 

As discussed before, foam targets can be used to generate near-critical ($n_0 \sim 0.1-1.0 n_c$) homogeneous plasmas. In the following, we therefore assume a near-critical plasma, i.e. a preionized and homogenized foam target. Furthermore, we assume an additional Gaussian channel at the optical axis to improve further laser guidance and retention of the strong laser fields.

\begin{figure}
    \centering
    \includegraphics[width=1\linewidth]{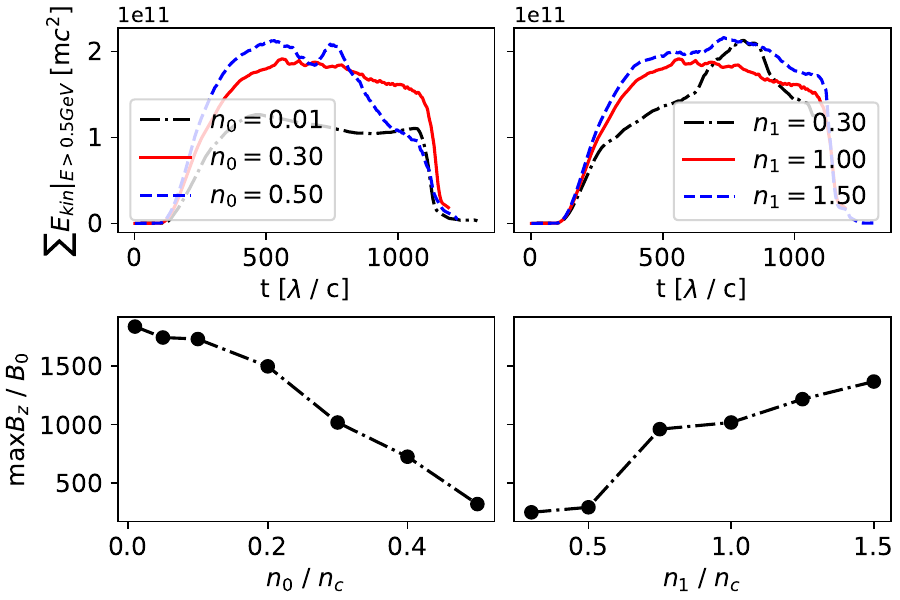}
    \caption{Total kinetic energy of high-energy ($E_\mathrm{kin} > 0.5$ GeV) electrons over time (top) for different plasma channel parameters and the maximum $B_z$ field for different parameters. Results from a 2D simulation with $n_0 = 0.3 n_c$ , $n_1 = n_c$ and $\sigma = 10\lambda$ unless stated otherwise.}
    \label{fig:dla-paramter}
\end{figure}

In order to generate a high number of pairs, electrons with high energy are required. In the proposed setup, electrons from the foam plasma are accelerated via direct laser acceleration. This can be easily seen from individual particle trajectories of high-energy electrons as plotted in Fig. \ref{fig:traj} which display clear characteristics of DLA (see Refs. \cite{Pukhov1999-gd, Arefiev2016-kb} for detailed discussions of DLA trajectories). 
The plasma parameters have to be chosen in such a way that both high-energy electrons and high fields can be achieved. While low inner densities $n_0 \lesssim 0.2 n_c$ lead to the highest amplification of the laser field in the reflector, the total energy gain of the electrons is considerably lower (cf. Fig. \ref{fig:dla-paramter}). Conversely, for high densities $n_0 \gtrsim 0.3 n_c$, the laser pulse quickly depletes, and achievable amplification quickly drops.
The impact of the outer density $n_1$ is found to be largely negligible on the acceleration and laser guidance in our simulations. 
While a tighter channel leads to a more rapid acceleration of the electrons at the start, the overall energy is largely unaffected. However, both too tight ($\sigma_C \lesssim 5\lambda$) or too wide ($\sigma_C \gtrsim 10\lambda$) channels can lead to significantly decreased field amplification at the reflector.
Consequently, it is non-trivial to find a parameter configuration such that the electrons are accelerated sufficiently while maintaining a strong laser field.

\begin{figure}
    \centering
    \includegraphics[width=0.95\linewidth]{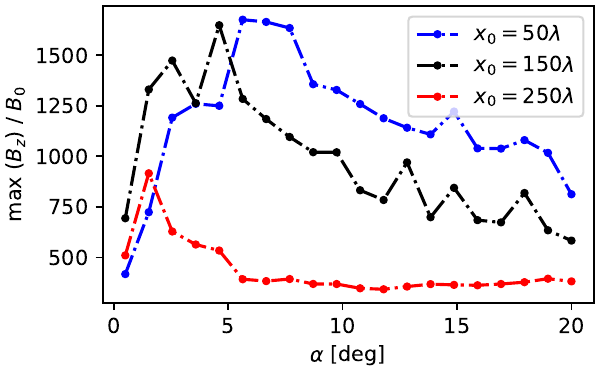}
    \caption{Amplification of the field amplitude for different opening angles $\alpha$ and positions $x_0$ of the reflector target.}
    \label{fig:opening_angle}
\end{figure}

Beyond the plasma channel structure, the reflected field amplitude can be further amplified by choosing a well-suited reflector geometry. Here we opt for a cone-shaped reflector with some opening angle $\alpha$. This focuses the incident laser pulse into an ever smaller cross-section, resulting in a significant increase in the field amplitude. Varying the opening angle, we find that the strongest $B_z$ is generated for an angle around $5^\circ$ for a reflector that is positioned at $150 \lambda$ (cf. Fig. \ref{fig:opening_angle}).  For a diverging laser pulse, the optimal opening angle is dependent on the position of the reflector target, making the optimal choice of an opening angle non-trivial.

The final field amplitude is strongly dependent on the length of the foam target: while a certain acceleration distance is required to attain high electron energies via DLA, the prolonged interaction of the laser pulse with the plasma can degrade its quality significantly.
Our simulations show that a short acceleration stage with moderate electron energies ($\gamma \sim 2000$) is already sufficient to create a large number of pairs. In fact, the simulations with a shorter interaction volume ($L \lesssim 200 \lambda$) yield the highest number of pairs.
This indicates that a better laser beam quality and, therefore, higher amplification at the interaction point is more important than an increase in electron energy via prolonged DLA.
However, the correlation is not perfect, as between $L = 200\lambda$ and $L=500\lambda$ the electron energy increases significantly, while achieving similar or even greater acceleration at the reflector, but producing fewer pairs overall.

\begin{figure}
    \centering
    \includegraphics[width=0.95\linewidth]{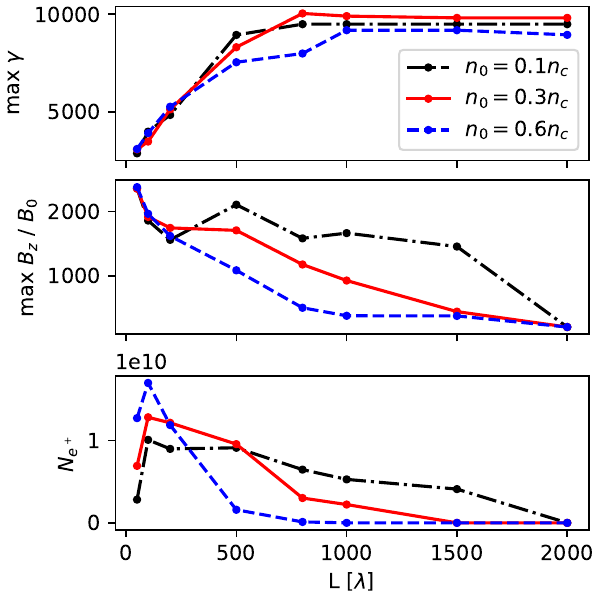}
    \caption{Maximum $\gamma_{e^-}$ (top), maximum $B_z$ (middle) and total pairs created (bottom) against the length of the foam pre-plasma, from 2D simulations.}
    \label{fig:distance_dep}
\end{figure}

For an ultra-short interaction volume with $L \lesssim 100 \lambda$, where the foam plasma exists solely inside the cone target, the laser pulse remains stable, leading to a significant amplification of the fields. Fig. \ref{fig:distance_dep} suggests that a significantly higher pair yield can be expected in this situation.

\begin{figure}
    \centering
    \includegraphics[width=1\linewidth]{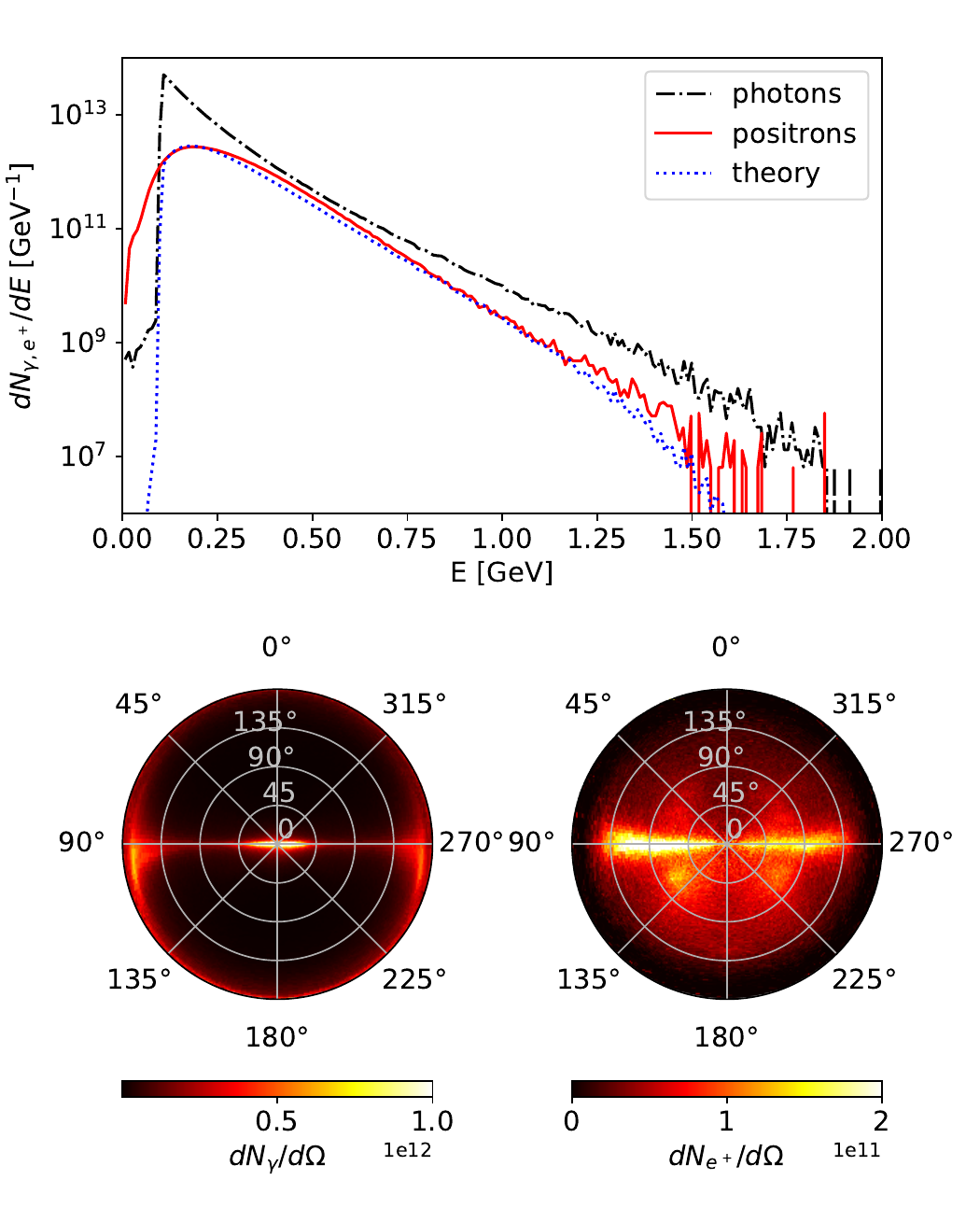}
    \caption{Energy spectra (top) of all generated photons (black) and positrons (red), as well as the angular distribution of photon (bottom left) and positron (bottom right) momenta. Photons with energy $E < 100$ MeV get omitted. The azimuthal angle $\varphi \in [0,2\pi]$ is relative to the laser-polarisation axis $\mathbf{e}_z$ and the polar angle $\theta \in [0,\pi]$ is in relation to the laser propagation axis $\mathbf{e}_x$.  }
    \label{fig:spectra}
\end{figure}

We conduct a full 3D simulation, where electrons are accelerated along a $50 \lambda$ homogenous foam target located entirely inside the cone cavity region. The plasma is homogeneous with density $n_0 = 0.3 n_c$, and we choose a reflector opening angle of $\alpha \sim 12^\circ$. For these parameters, we observe a maximum field strength of $B_z /B_0 \sim 2000$ and a maximum electron energy $\sim 2.5$ GeV.
While this short distance is not sufficient to achieve the maximum electron energy, due to the advantageous reflection, we achieve $\chi \sim 10 \gg 1$, facilitating pair production outside of the exponentially damped regime.
A large number of thermal photons get created at the interaction point (see Fig. \ref{fig:spectra}), with a maximum energy of $\sim 2$ GeV. The generated positrons exhibit an energy peak at $E_\mathrm{max} \sim 0.2$ GeV, with a long falling flank up to $\sim 1.75$ GeV. Comparing the obtained positron spectrum with the theoretical prediction (cf. Eq. \eqref{eq:rate_positron}) shows good qualitative agreement when weighted by the photon number. 

\begin{figure}
    \centering
    \includegraphics[width=1\linewidth]{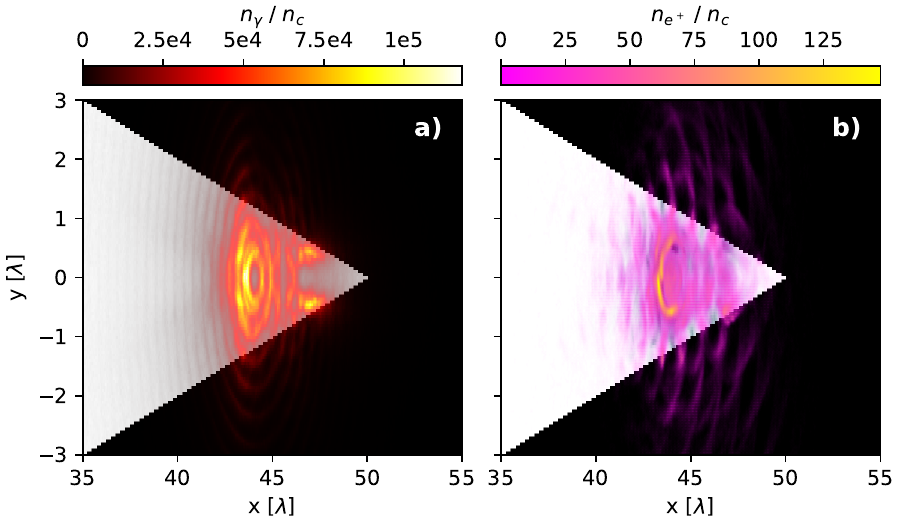}
    \caption{Photon density (a) and positron density (b) at T=160$\lambda / c$.}
    \label{fig:ultrashort}
\end{figure}

Photons are mainly emitted in the forward direction and perpendicular to the polarisation axis of the laser.
At the interaction point, where the fields get amplified to many times their initial amplitude, a high number of photons are created, clearly showing a ring-like structure, corresponding to the periodicity of the laser pulse. Due to their subsequent decay, a high-density electron-positron plasma emerges, far exceeding the initial electron density $n_{e^+} \gg n_0$, showing a similar periodic pattern (cf. Fig. \ref{fig:ultrashort}).
In total, we generate up to $\sim 10^{11}$ electron-positron pairs with a local peak density $n_{e^+} \sim 30 n_c$.

While the laser pulses in PIC simulations are usually considered to be of Gaussian shape, the existence of pre-pulse(s) and pedestal can strongly influence the laser-matter interaction that occurs when the main peak irradiates the target. Specifically for ion acceleration, this can have drastic consequences \cite{Boller2025}.

As shown, e.g. in Ref. \cite{Howard2025}, the temporal profile of high-intensity laser pulses exhibits strong ``wings''. While in the case of ATLAS-3000, the intensity of such wings is on the order of $\sim 10\%$, if we consider a field amplitude of $a_0 \sim 200$ in our paper, a comparably strong wing would already have an intensity of $a_0 \sim 60$. Especially since this wing precedes the main pulse by multiple hundreds of femtoseconds, the interaction of the laser pulse with the foam target will be modulated quite strongly.

Thus, we conduct further simulations to investigate pre-pulse effects. In a first run, we consider a non-homogenised foam to see if the foam can be fully homogenised. We consider a weak prepulse with amplitude $a_0 =0.2$ and duration $1$ ps that preceeds the main pulse by multiple picoseconds. The simulations indicate the pulse is intense enough to fully homogenise the target such that our assumptions of a homogenous, fully ionised plasma, for the previous simulation runs are warranted.

Moreover, in a separate simulation, we study the effect of the laser wings: we model one such wing as a separate laser pulse with $a_0 = 60$  and duration $40$ fs that precedes our main pulse by $\sim 300$ fs. 
The simulations show that the preceding wing is already sufficient to almost completely evacuate the interaction domain of electrons, such that the main laser pulse is no longer able to efficiently accelerate them and create an abundance of pairs. When considering such a setup, the resulting number of generated pairs is multiple orders of magnitude (roughly a factor of 1000) smaller. However, the number of generated pairs is still significantly larger compared to only a laser pulse of $a_0 \sim 60$, indicating that the main laser pulse still contributes to the pair generation process.

\section{Conclusion}
Using PIC simulations, we have investigated a single-laser setup for probing the strong-field quantum electrodynamic regime $\chi \gtrsim 1$, where nonlinear inverse Compton scattering and nonlinear Breit-Wheeler pair creation become relevant.
The laser pulse drives electrons from a homogenised foam target to high energies via direct laser acceleration (DLA). Subsequently, the laser is reflected by a cone-shaped target placed at the end of the foam, leading to the interaction of the GeV electrons with the reflected laser field and, finally, the creation of electron-positron pairs.

The length of the foam target affects the attainable electron energies, while the presence of a channel structure and the opening angle of the cone reflector influence the collimation of the laser beam during its propagation.

We find that the strong amplification of the reflected field is more important than the electron energies achieved via DLA. An optimised setup is shown to produce a total of $\sim 10^{11}$ electron-positron pairs using a 1.5 kJ laser pulse. 

\begin{acknowledgments}
The authors would like to thank Ke Jiang (SZTU) for fruitful discussions. 
This work has been supported by BMBF (Project 05P24PF1).
The authors gratefully acknowledge the Gauss Centre for Supercomputing e.V. \cite{GCS} for funding this project (lpqed) by providing computing time through the John von Neumann Institute for Computing (NIC) on the GCS Supercomputer JUWELS at J\"ulich Supercomputing Centre (JSC). 
\end{acknowledgments}

\bibliographystyle{apsrev4-2}
\bibliography{bib.bib}

\end{document}